\documentclass[12pt]{iopart}
\usepackage{graphicx}
\usepackage{multirow}
\usepackage{dcolumn}
\begin{document}

\title{High-precision spectroscopy of ultracold molecules in an optical lattice}

\author{B. H. McGuyer$^1$, M. McDonald$^1$, G. Z. Iwata$^1$, M. G. Tarallo$^1$, A. T. Grier$^1$, F. Apfelbeck$^{1,2}$, and T. Zelevinsky$^{1,*}$}

\address{$^1$ Department of Physics, Columbia University, 538 West 120th Street, New York, New York 10027-5255, USA}
\address{$^2$ Faculty of Physics, Ludwig Maximilian University of Munich, Schellingstrasse 4, 80799 Munich, Germany}
\vspace{10pt}

\begin{abstract}
The study of ultracold molecules tightly trapped in an optical lattice can expand the frontier of precision measurement and spectroscopy, and provide a deeper insight into molecular and fundamental physics.  Here we create, probe, and image microkelvin $^{88}$Sr$_2$ molecules in a lattice, and demonstrate precise measurements of molecular parameters as well as coherent control of molecular quantum states using optical fields.  We discuss the sensitivity of the system to dimensional effects, a new bound-to-continuum spectroscopy technique for highly accurate binding energy measurements, and prospects for new physics with this rich experimental system.
\end{abstract}

\pacs{37.10.Jk, 34.80.Qb, 82.53.Kp, 33.80.Gj}
%

\vspace{2pc}
\noindent{\it Keywords}: Ultracold Molecules, Photoassociation, Photodissociation, Optical Lattice, Molecular Clock.
\newline\noindent
%
%
%


$^*$ Email:  tz@phys.columbia.edu
\section{Introduction}
Many of the concepts and methods that emerged from research with ultracold atoms are gaining traction with more complex physical systems.  In this work, we demonstrate a combination of light-assisted molecule formation \cite{JonesRMP06} with optical lattice clock techniques \cite{KatoriNPhot11_LatticeClocks,LudlowHinkleyScience13_10to18YbComparison,YeBloomNature14_10to18SrComparison} that yields optical and microwave spectra of molecules with unprecedented resolution.
Using a narrow optical intercombination transition, we create, probe, and image diatomic strontium molecules, $^{88}$Sr$_2$, in an optical lattice.
We observe signatures of trap dimensionality in the photoassociation (PA) spectra of the ultracold colliding atoms that have been predicted but not yet reported \cite{julienne:2006}.  These spectra allow a precise determination of molecular transition strengths.  The transition strength measurements are in good agreement with a state-of-the-art {\it ab initio} quantum chemistry model \cite{MoszynskiSkomorowskiJCP12_Sr2Dynamics}, and are essential for designing efficient molecule formation pathways, as well as for refining the toolkit of {\it ab initio} molecular calculations.
We describe a new method of measuring molecular binding energies using photodissociation (PD) spectra, involving transitions between bound and continuum states of the atoms.
This method, which is also sensitive to the dimensionality of the trapping potential, yields precise binding energies with kilohertz uncertainties that are essential for rigorous studies of molecular quantum electrodynamical effects as a function of interatomic separation.
Furthermore, we achieve coherent control of molecular quantum states, which is a prerequisite for a `molecular lattice clock', a competitive tool for metrology and precision measurements that is sensitive to different physics than atomic microwave or optical clocks \cite{ZelevinskyPRL08}.

The $^{88}$Sr isotope is advantageous for this work due to its electronic structure, which is characteristic of alkaline-earth-like elements \cite{stellmer:2014}.  Direct laser cooling of Sr yields sub-$\mu$K temperatures \cite{KatoriPRL99}.
Furthermore, the availability of optically accessible singlet and triplet electronic states allows state-insensitive lattice trapping \cite{YeSci08}, and provides multiple narrow optical `clock' transitions that facilitate the manipulation of atomic and molecular quantum states.
While the lack of hyperfine structure precludes magnetoassociation of Sr \cite{KohlerRMP06}, narrow-line photoassociation is highly effective \cite{zelevinsky:2006}.  The spinless $^1S_0$ ground state ensures a simple interatomic interaction potential, which together with advanced data and models available for this atom \cite{TiemannSteinEPJD10_Sr2XPotential,TiemannSteinEPJD11_Sr2ExcPotentials,MoszynskiSkomorowskiJCP12_Sr2Dynamics} makes it an ideal candidate for several classes of precision measurements, as discussed below.

This Article is organized as follows.
In Section \ref{sec:PA}, we describe high-resolution PA of $^{88}$Sr atoms in an optical lattice, which we use to study the effects of lattice confinement on atomic collisions and to measure the strengths of molecular transitions.
We discuss how nonadiabatic Coriolis mixing affects the determination of Franck-Condon factors (FCFs) from two-color PA spectra.
In Section \ref{sec:MoleculeSpectroscImaging}, we describe how the transition strength results lead to efficient ultracold $^{88}$Sr$_2$ molecule production and detection, with an emphasis on converting molecules back to atoms for convenient absorption imaging.
We also demonstrate how molecular PD can yield precise absolute binding energies.
Section \ref{sec:CoherentControl} presents coherent control of molecular quantum states in the lattice.
Finally, in Section \ref{sec:Outlook} we provide a glimpse of precision measurements in molecular and fundamental physics that can benefit from the refined spectral resolution and coherent control of molecules achieved in this work.


\section{High-resolution photoassociation spectroscopy}
\label{sec:PA}
To obtain ultracold, reasonably dense samples of Sr$_2$ molecules, we initially laser cool $^{88}$Sr atoms to temperatures $T\sim2$ $\mu$K in a one-dimensional (1D) optical lattice \cite{reinaudi:2012}.
The lattice is formed by a retroreflected beam of linearly polarized light and operated near the magic wavelength for the $^1S_0-{^3P_1}$ atomic intercombination line, $\sim914$ nm \cite{ido:2003}.
The trapped atoms are manipulated with narrow-linewidth ($<200$ Hz) laser light (689 nm wavelength) along the tight-confinement axis in the Lamb-Dicke and resolved-sideband regimes \cite{leibfried:2003}.
This laser light can be tuned to perform either one- or two-color PA to probe electronically excited or ground vibrational states, respectively, or to populate specific rovibrational states to serve as a starting point for further experiments.
Afterwards, atoms that either remain in the trap or are recovered by dissociating the molecules are counted via absorption imaging using blue light (461 nm wavelength) resonant with the strong atomic $^1S_0-{^1P_1}$ line.

\begin{figure}[t]
	\flushright
	\includegraphics[]{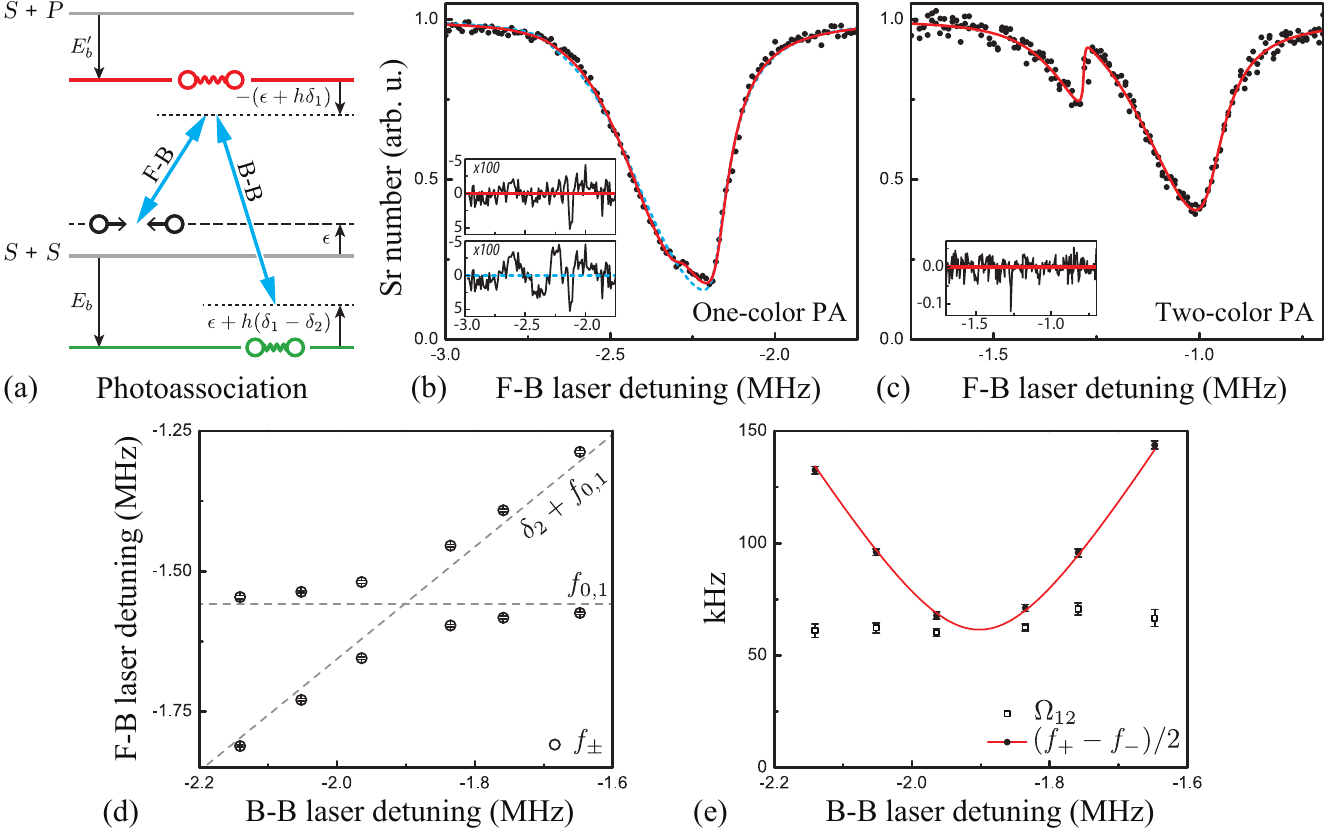}
	\caption{
Intercombination-line photoassociation (PA).
(a) Energy level diagram and detuning conventions for free-bound (F-B) and bound-bound (B-B) PA lasers (blue arrows).
The F-B laser enables a collision of two ground-state atoms (black) with relative kinetic energy $\epsilon$ to couple to an excited-state molecule (red), which in turn may be coupled to a ground-state molecule (green) by the B-B laser.
(b) One-color PA with the F-B laser. 
The spectrum is fit with Eq. (\ref{Ntau}).
The solid red curve is a fit using the probability function (\ref{eq:PDimensional}) with residuals shown in the top inset, and the dashed blue curve uses the probability (\ref{P1}) with residuals in the bottom inset.
(c) Two-color PA with both lasers.
The B-B laser introduces a visible Autler-Townes splitting of the line shape, described by Eq. (\ref{P2}) with residuals shown in the inset.
(d) Example resonance positions $f_\pm$ of Eq. (\ref{fpm}) from fitting two-color PA spectra.
(e) The Rabi frequency $\Omega_{12}$ is determined from two-color PA line fits either directly or by fitting the spacings $(f_+ - f_-)/2$ with a hyperbola.
}
	\label{fig1}
\end{figure}
Figure \ref{fig1}(a) schematically illustrates one- and two-color PA and defines the free-bound (F-B) and bound-bound (B-B) laser frequency conventions, while Figures \ref{fig1}(b,c) demonstrate high-resolution PA spectra near the intercombination atomic line.
The PA resonances are only a few hundred kilohertz wide, and are sensitive to the quasi-2D nature of ultracold atomic collisions in the lattice.
Following Ref. \cite{zelevinsky:2006}, the 1D-lattice PA spectra are described by the line shape
\begin{eqnarray}	\label{Ntau}
N(\tau) = \frac{N_0}{ 1 + A \int_0^\infty p(\epsilon) \, e^{-\epsilon/(k_B T)} \, d\epsilon/(k_B T)},
\end{eqnarray}
which assumes that one-body losses are negligible, and involves integration over the collision energy $\epsilon$.
Here, $T$ is the initial atomic temperature, $k_B$ is the Boltzmann constant, $N(\tau)$ is the number of Sr atoms after a PA pulse of duration $\tau$, and $N_0=N(0)$ is the number of atoms remaining in the trap far from a PA resonance.
The denominator in Eq. (\ref{Ntau}) is equivalent to $1 + 2 K n_0 \tau$, where $n_0 = n(0)$ is the initial atomic number density and $K$ is the conventional PA rate satisfying $dn/dt = - 2 K n^2$.
As written, the dimensionless fit parameter
\begin{eqnarray}
A = 2 h l_{\rm opt} n_0 \tau / \mu,
\end{eqnarray}
where $h=2\pi\hbar$ is the Planck constant and $\mu$ is the two-body reduced mass,
depends on the optical length $l_{\rm opt}$ parameterizing the F-B transition strength,
which is independent of $\epsilon$ for ultracold collisions \cite{zelevinsky:2006,LoptIsConstant}.
Note that the optical lattice critically affects the relationship between the experimentally accessible number $N$ and the local number density $n$ \cite{borkowski:2014,osborn:thesis}, which must be accounted for in using $A$ to determine $l_{\rm opt}$ or vice versa.
In this work, we left $A$ as a free parameter.
For quasi-2D collisions in a 1D optical lattice, the probability function
\begin{eqnarray}
\label{Pdef}
p(\epsilon) = |S|^2 \, \gamma_1/\gamma_s
\end{eqnarray}
is given by a scattering-matrix probability $|S|^2$
and the ratio of natural
($\gamma_1\sim15$ kHz \cite{mcguyer:1g}) 
and stimulated decay rates.
For one-color $s$-wave PA \cite{zelevinsky:2006,julienne:1996}, this function is
\begin{eqnarray}	\label{P1}
p_1(\epsilon) = \frac{\gamma_1^2}{(\epsilon/h + \delta_1)^2 + (\gamma_1 + \gamma_b + \gamma_s)^2/4},
\end{eqnarray}
where the detuning $\delta_1$ follows the convention in Figure \ref{fig1}(a).
For typical PA laser powers where $\gamma_s \ll \gamma_1$, we introduce an $\epsilon$-independent broadening parameter $\gamma_b$ in $p(\epsilon)$ here and below to account for the observed spectral widths following Ref. \cite{zelevinsky:2006}.

The PA spectra in Figure \ref{fig1} are sensitive to additional dimensional effects from the lattice, which strongly quantizes the axial motion of the colliding atoms.
As demonstrated by the residuals in the inset of Figure \ref{fig1}(b), the spectra are best described by taking into account the thermal distribution of axial trap states of the colliding atoms and, in particular, the collisions between atoms in different axial trap states.
For the quasi-2D regime, parity considerations restrict interactions to be between atoms in axial trap states separated by even multiples of the trap energy quantum, $\hbar\omega_x$, where $\omega_x$ is the axial angular frequency of the optical lattice.  Hence, the relative axial collision energies are spaced by $2\hbar\omega_{x}$.
This additional lattice-quantized collisional physics is included by substituting
\begin{eqnarray}
\label{eq:PDimensional}
p_{\rm coll}(\epsilon) = \sum_{n=0}^\infty e^{-2 n \hbar \omega_x/(k_B T)} \, p_1(\epsilon + 2 n \hbar \omega_x)
\end{eqnarray}
for $p(\epsilon)$ in Eq. (\ref{Ntau}) for one-color PA \cite{julienne:2006,petrov:2001}.
The signature `hump' of this collisional energy quantization is observed in our system because of the tight lattice trapping and of the inherently narrow PA spectra due to long atomic lifetimes and ultracold temperatures.
Related effects have been observed in radio-frequency (RF) dissociation of $^6$Li$_2$ Feshbach molecules in an optical dipole trap \cite{zurn:2013}.

If during PA a second laser is added which is nearly resonant with a transition between the excited level being probed and a ground-state rovibrational level of the molecule, the PA resonance is split into an Autler-Townes doublet \cite{TownesAutlerPR55_StarkEffectACFields}.
For such two-color $s$-wave PA \cite{julienne:1996}, the function (\ref{Pdef}) is
\begin{eqnarray}	\label{P2}
\fl
\quad
p_2(\epsilon) = \frac{\gamma_1^2 (\epsilon/h + \delta_1 - \delta_{2})^2}{[\Omega_{12}^2 - (\epsilon/h + \delta_1)(\epsilon/h + \delta_1 - \delta_{2})]^2 + (\epsilon/h + \delta_1 - \delta_{2})^2(\gamma_1 + \gamma_b + \gamma_s)^2/4},
\end{eqnarray}
where $\Omega_{\rm 12}$ is the bound-bound Rabi frequency.
The detunings $\delta_1$ and $\delta_2$ follow the conventions in Figure 1(a), and relate to the F-B and B-B laser frequencies as $\delta_1 = f_{\rm FB} - f_{0,1}$ and $\delta_2 = f_{\rm BB} - f_{0,2}$, where the fit parameters $f_{0,1}$ and $f_{0,2}$ represent on-resonance laser frequencies.  Figure \ref{fig1}(c) shows a typical fitted two-color PA line shape, with Autler-Townes peaks near the two F-B laser frequencies
\begin{eqnarray}
\label{fpm}
f_\pm = f_{0,1} + \delta_2/2 \pm \sqrt{\delta_2^2/4 + \Omega_{12}^2}.
\end{eqnarray}
The peak positions follow an avoided crossing as the B-B laser frequency is varied (Figure \ref{fig1}(d)).

\begin{table}
\caption{\label{tab:1}Molecular transition strengths measured via two-color PA that are relevant to ultracold molecule creation.
The values are the absorption oscillator strengths for the indicated transitions, normalized to the chosen transition with unity strength.
Experimental uncertainties are indicated, and {\it ab initio} theoretical values \cite{MoszynskiSkomorowskiJCP12_Sr2Dynamics,borkowski:2014,SkomorowskiPrivate} are shown beneath the measured values.
Binding energies are given to the nearest MHz \cite{reinaudi:2012,killian:2008,zelevinsky:2006}.
The state assignment of the two deepest excited levels follows \cite{zelevinsky:2006,borkowski:2014}.
}
\begin{indented}
\item[]\begin{tabular}{@{}r r r l l l l l}
\br
\centre{2}{X$^1\Sigma_g^+$} 	& 		& \centre{4}{$(1)0_u^+$}			& $(1)1_u$ \\
\ns \ns
	& 						& 		& \crule{4} 						& \crule{1} \\
\ns
	& 						& $v'$: 	& $-3$ 	& $-4$ 	& $-5$ 	& $-6$ 	& $-3$ \\	
$v$ 	& $E_b$		 			& $E_b'$: 	& 222 	& 1084 	& 3463 	& 8429 	& 8200 \\
\mr
$-1$	& 137  					& 		& 0.66(8)	& 0.53(2)	& 		& 		& 	\\
	&						& 		& 0.606	& 0.545	& 		& 		& 	\\
$-2$	& 1400 					& 		& 		& 0.158(6)	 & 1.00(8)	& 		& 	\\
	&						& 		& 		& 0.144	& 1		& 		& 	\\
$-3$	& 5111 					& 		& 		& 		& 		& 1.88(9)	& 0.46(3)	\\
	&						& 		& 		& 		& 		& 1.651	& 0.553	\\
\br
\end{tabular}
\end{indented}
\end{table}
By finding the B-B laser frequencies that lead to minimum frequency separations of the Autler-Townes doublets,
as shown in Figure \ref{fig1}(e),
along with B-B laser intensities for different pairs of ground and excited states, molecular transition strengths can be determined.
Table \ref{tab:1} shows the transition strengths we have measured from such two-color PA spectra for several pairs of the most weakly bound molecular levels within the ground electronic state X$^1\Sigma_g^+$ and excited electronic states $(1)0_u^+$ and $(1)1_u$.  The negative signs of the ground- and excited-state vibrational quantum numbers, $v$ and $v'$, imply counting down from the threshold.  The ground- and excited-state total angular momenta and their projections onto the laboratory axis are $J=0$, $J'=1$ and $m=m'=0$.
The table presents relative strengths, which are more precise than absolute strengths because of the difficulty in accurately determining the local intensity of the focused probe laser beams.
The strengths are proportional to $\Omega_{12}^2/P$ \cite{mcguyer:1g}, where $P$ is the F-B laser power.  Theoretical values \cite{SkomorowskiPrivate} agree well with the measurements, in all cases matching experiment within $3\times$ the tight experimental uncertainty, confirming the validity of the {\it ab initio} molecular model \cite{MoszynskiSkomorowskiJCP12_Sr2Dynamics}.  The transitions that were experimentally verified are the ones with the largest expected strengths, which are most relevant to molecule creation.

Note that dimensionless vibrational wave function overlaps, or Franck-Condon factors (FCFs) $f_{v,v'} = | \langle v | v' \rangle|^2$, are commonly used to quantify transition strengths near threshold.  Here, however, nonadiabatic Coriolis mixing of the $0_u^+$ and $1_u$ potentials \cite{mcguyer:2013} leads to interference effects in transitions to the excited states near the intercombination-line asymptote.  Neglecting this mixing, measurements could determine $f_{v,v'} = \Omega_{12}^2/ (2 \alpha^2 \Omega_a^2)$ via the ratio of the molecular and atomic Rabi frequencies, $\Omega_{12}$ and $\Omega_a$ \cite{reinaudi:2012},
where $\alpha^2 = 1/3,\;2/3$ for transitions to $0_u^+$, $1_u$ \cite{julienne:2001}.
Alternatively, using the theoretical prediction of $f_{-2,-5} = 0.508$ with this relationship, the values in Table 1 can be converted to estimates of FCFs.
This yields good agreement for all but the two deepest excited states, where for example $f_{-3,-6} \approx 1.51$, which is unphysical because it exceeds unity.
The procedure fails because the actual molecular state $0_{\rm u}^+(v'=-6)$ has both $0_u^+$ and $1_u$ parts due to Coriolis mixing, and the contributions from these parts constructively interfere in this transition.

\section{Ultracold molecule spectroscopy and imaging}
\label{sec:MoleculeSpectroscImaging}
The relative transition strength measurements presented in Table \ref{tab:1} contain several values of order unity, which correspond to FCFs of roughly the same magnitude.  This implies the existence of highly efficient molecule creation pathways via one-color PA followed by well-directed spontaneous decay \cite{reinaudi:2012}.
Specifically, PA to $0_u^+(v'=-5,J'=1)$ efficiently produces X$^1\Sigma_g^+(v=-2,J)$, and PA to
$0_u^+(v'=-3,J'=1)$ or to $0_u^+(v'=-4,J'=1)$ produces X$^1\Sigma_g^+(v=-1,J)$, both with $J=0,2$.
(Note that for symmetry reasons, the ground-state potential only supports even values of $J$.)
This procedure also produces a small number of molecules with higher values of $J$, likely because of incoherent scattering of the PA laser by the newly produced molecules.
Note that while $0_u^+(v'=-6)$ and $1_u(v'=-3)$ have high transition strengths to X$^1\Sigma_g^+(v=-3)$, experimentally these states do not yield large numbers of ground-state molecules.  This is likely because the ground-state molecule production efficiency depends also on $l_{\rm{opt}}$, which rapidly decreases with increased excited-state binding energy, resulting in PA rates that are too slow compared to incoherent scattering rates.

\begin{figure}[t]
	\flushright
	\includegraphics[]{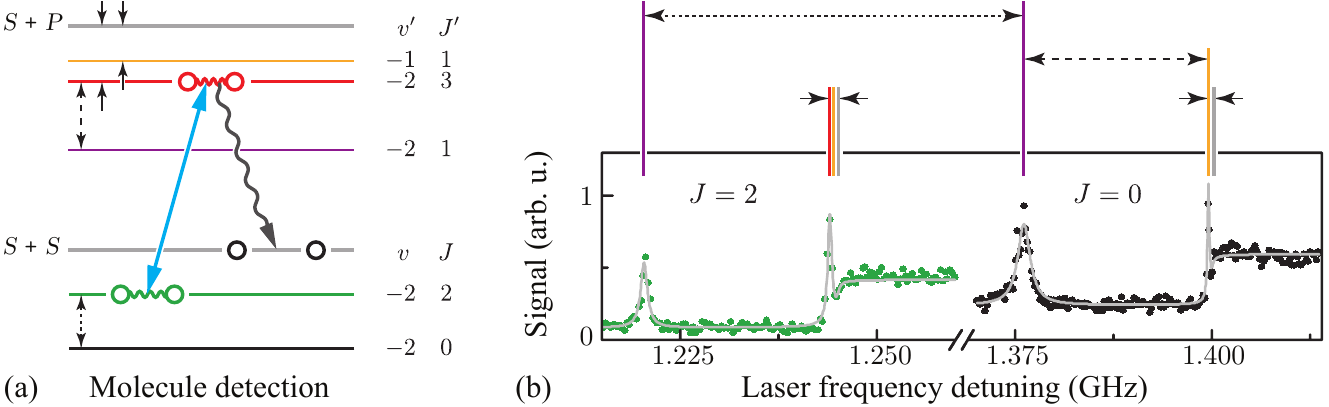}
	\caption{
Molecule detection scheme.
(a) Energy level diagram (not to scale) illustrating how a probe laser (blue arrow) may drive ground-state Sr$_2$ molecules through an atom-recovery transition to an excited state $0_u^+(v',J')$ that spontaneously decays to unbound Sr atoms.
(b) Sample spectra produced with these recovery transitions for X$^1\Sigma_g^+(v=-2)$ molecules with $J=0$ (black) and $J=2$ (green).
Besides peaks corresponding to bound-bound transitions, the spectra show bound-free `shelf' line shapes for transitions into the excited-state continuum that fragment the molecules \cite{mcguyer:2013}.
The arrows in both panels highlight how we are able to precisely and accurately measure binding energies or rotational spacings using atom-recovery spectra.
}
	\label{fig2}
\end{figure}
Figure \ref{fig2} illustrates detection mechanisms for the resulting ground-state molecules \cite{reinaudi:2012,mcguyer:2013}.
Using `atom-recovery' transitions, we convert the molecules to ground-state Sr atoms that are counted by absorption imaging.  This is accomplished by either transferring the molecules to excited states that spontaneously decay to Sr atoms with a high probability, or by directly fragmenting (or photodissociating) the molecules using the excited-state atomic continuum.
Because of nonadiabatic Coriolis mixing, these recovery transitions feature anomalously large Zeeman shifts that readily enable the selective detection of ground-state magnetic sublevel populations \cite{mcguyer:2013}.
Interestingly, the Coriolis mixing is also responsible for the existence of the least-bound $J'=3$ state shown, which is predicted to be unbound by calculations that do not include this mixing \cite{SkomorowskiPrivate}.
As a result, this state is expected to be highly sensitive to long-range physics.

Besides the electric-dipole transitions shown, there are also magnetic-dipole and electric-quadrupole transitions that are efficient for detecting X$^1\Sigma_g^+(v=-1)$ molecules.
These transitions are enabled by the least-bound vibrational level of the gerade $(1)1_g$ potential, which has a $J'=1$ state with binding energy $E_b' = 19.0$ MHz, and a $J'=2$ state with $E_b' = 7.2$ MHz \cite{mcguyer:1g}.
Because of the subradiant nature of the $1_g$ potential, many of its bound states provide extremely narrow molecular transitions.

\begin{table}
\caption{\label{tab:2}
Energies measured by the atom-recovery transitions of Figure 2.
The binding energies $E_b'$ are of $(1)0_{\rm u}^+(v',J')$ excited states, and $E_b$ of X$^1\Sigma_{\rm g}^+(v,J)$ ground states.
Values in parenthesis are uncertainties on the last digits and include extrapolating to zero lattice power, probe power, and magnetic field.
The values of $E_b$ from this work are roughly an order of magnitude more precise than previously measured values \cite{zelevinsky:2006}.
}
\begin{indented}
\item[]\begin{tabular}{@{}llll}
\br
Energy (MHz)		& This work		& Previous work \cite{zelevinsky:2006} \\
\mr
$E_b'(v'=-1,J'=1)$ 	& 0.4521(43) 	& 0.435(37) \\
$E_b'(v'=-2,J'=1)$ 	& 23.9684(50)  	& 23.932(33) \\
$E_b'(v'=-2,J'=3)$ 	& 0.626(12)  	&  \\
$E_b(v=-2,J=2) - E_b(v=-2,J=0)$ 	& 154.5497(46) \\
\br
\end{tabular}
\end{indented}
\label{tab2}
\end{table}
By measuring energy spacings between atom-recovery transitions using excited molecular or continuum states, we precisely determined the excited-state binding energies and rotational splittings given in Table \ref{tab:2}.
To determine absolute binding energies, we use bound-free recovery transitions with `shelf' line shapes as frequency references with $\sim2$ kHz uncertainty per trace.  The values quoted in Table \ref{tab:2} have been extrapolated to zero lattice power, probe power, and magnetic field to remove light and Zeeman shifts.

\begin{figure}[t]
	\flushright
	\includegraphics[]{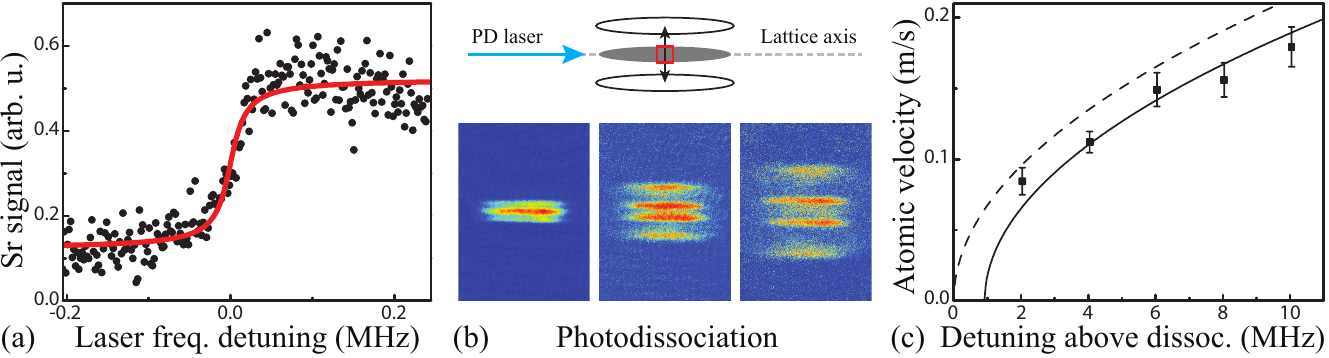}
	\caption{
Intercombination-line molecular photodissociation (PD).
(a) A `shelf' line shape obtained by optically exciting ground-state X$^1\Sigma_g^+(v=-2,J=0)$ molecules to the excited-state $^1S_0+{^3P_1}$ atomic continuum.
The spectrum is fit with Eq. (\ref{eq:W2D}).
The slight loss of signal with increased detuning is due to atoms escaping the detection area [red box in part (b)].
(b) With increased detuning above the continuum, the resulting atoms have enough energy to escape the optical lattice.
Viewed off-axis as sketched (top) and observed with absorption imaging (bottom), this occurs mainly along a preferred direction set by the PD laser polarization.
Two pairs of vertically split clouds appear in the successively delayed absorption images:  an outer pair of faster atoms from $J=2$ PD, and an inner pair of slower atoms from $J=0$ PD.
(c) The measured velocities of the atoms in the slower, inner clouds.
The dashed curve is a speed limit from energy conservation.
The solid curve is a fit to the data with Eq. (\ref{eq:CloudSpeed}), offset by the lattice depth.
}
	\label{fig3}
\end{figure}

To analyze the bound-free `shelf' line shapes in spectra such as in Figure \ref{fig3}(a), we fit the spectra with the function
\begin{eqnarray}
W_{\rm 2D}(f) = W_0 + B \left\{ \arctan \left[ \frac{2(f-f_0)}{\gamma} \right] + \frac{\pi}{2} \right\},
\label{eq:W2D}
\end{eqnarray}
where $W_0$ is a baseline signal, $B$ is a shelf amplitude parameter, $\gamma$ is a transition width, $f$ is the probe laser frequency, and $f_0$ is the shelf center.
We motivate this line shape by noting that this PD process resembles time-reversed PA, but with the initial thermal ensemble of collision energies $\epsilon$ replaced by a fixed dissociation energy $h(f-f_0)$.
As with one-color PA, the optical lattice affects the dimensionality of this process.
The line shape (\ref{eq:W2D}) follows from adapting the quasi-2D $s$-wave PA line shape \cite{zelevinsky:2006}, which yields
$[W_{\rm 2D}(f) - W_0] \propto \int_0^\infty L(f-f_0, \epsilon) d\epsilon$, where the Lorentzian
$L(\delta, \epsilon) = [\gamma/\pi]/[(\epsilon/h + \delta)^2 + \gamma^2/4]$ with full-width-at-half-maximum (FWHM) $\gamma$.
This derivation assumes sufficiently low PD laser powers such that $\gamma_s/\gamma\ll1$, which is supported by our spectra.
Alternatively, the 3D $s$-wave PA line shape \cite{zelevinsky:2006} would suggest a different line shape,
$[W_{\rm 3D}(f) - W_0] \propto \int_0^\infty L(f - f_0, \epsilon) \sqrt{\epsilon} \, d\epsilon \propto \sqrt{(f - f_0) + \sqrt{(f - f_0)^2 + \gamma^2/4}}$.  However, this function does not accurately describe our observations for dissociation energies within the lattice depth.  For the measurements presented in Table \ref{tab:2}, we have assumed that the sharp onset of the continuum in Figure \ref{fig3}(a) is unaffected at our level of precision by the small, long-range centrifugal barrier of the $0_u^+(J'=1)$ molecular potential.

The atomic dynamics following molecular PD is illustrated in Figure \ref{fig3}.  Figure \ref{fig3}(b) shows our imaging geometry where the camera is facing the lattice off-axis, and the atomic clouds split in the vertical direction set by the PD laser polarization. The outer (more energetic) clouds correspond to $J=2$ molecules, and the inner (less energetic) clouds to $J=0$ molecules.  Figure 3(c) shows the estimated velocities of the resulting Sr atoms in the imaged plane.
Classically, the total kinetic energy of the atom pair should be given by the frequency detuning $(f - f_0)$ of the PD laser relative to the continuum, $h (f - f_0) =2 \times (m_{\rm Sr} v^2/2)$, where $m_{\rm Sr}$ is the mass of a $^{88}$Sr atom.  This motivates the fit function
\begin{eqnarray}
v(f) =
\cases{C \sqrt{f - f_0}&for $f - f_0 > 0$\\
0&for $f - f_0 \le 0$\\}.
\label{eq:CloudSpeed}
\end{eqnarray}
From conservation of energy, the parameter $C$ should be $\sqrt{h / m_{\rm Sr}} \approx 6.738 \times10^{-5}$ m/$\sqrt{\rm s}$.
This case is shown by the dashed curve in Figure 3(c).
In practice, as indicated by the solid curve in Figure 3(c), the angular distribution of the fragments tends to decrease the fitted value of $C$.
In addition, the fit function should be shifted rightward roughly by the optical lattice depth.

\section{Coherent manipulation}
\label{sec:CoherentControl}

\begin{figure}[t]
	\flushright
	\includegraphics[]{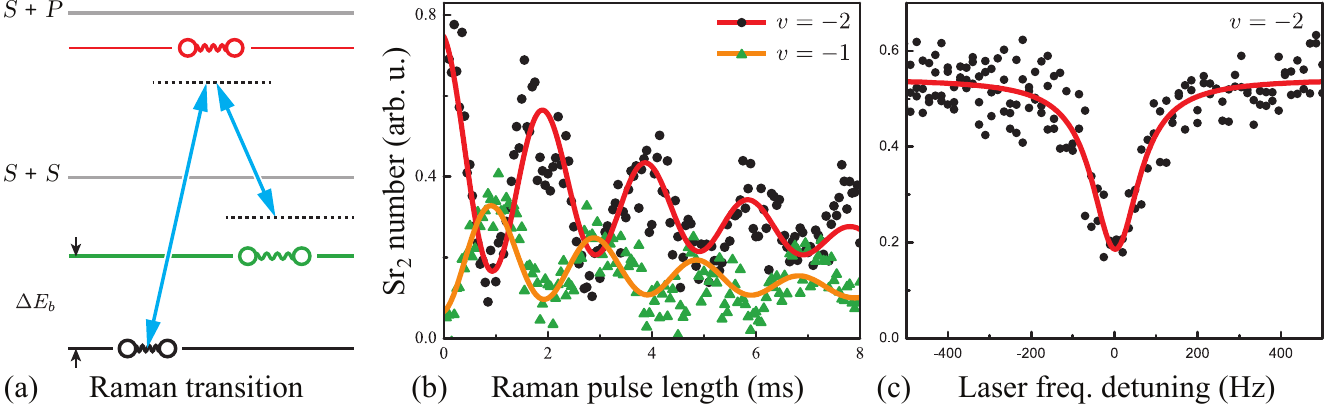}
	\caption{
Two-color Raman transitions between vibrational levels of ground-state Sr$_2$ molecules.
(a) Schematic energy level diagram.
(b) The detected populations of the initial and final vibrational levels with $J=0$ show coherent Rabi oscillations as a function of the Raman pulse length.
(c) The linewidths of the Raman transitions are mostly limited by thermal decoherence from differential lattice light shifts \cite{mcdonald:therm}.
For this trace, the experimental conditions were adjusted to achieve a narrow width, and a Lorentzian fit to the natural logarithm of the data (to account for linear probe absorption) yields a 112(8) Hz FWHM.
}
	\label{fig4}
\end{figure}
If the lifetime of a transition under investigation is longer than the probe pulse time, coherent oscillations between the initial and final states can be achieved provided that other sources of decoherence are minimized.  Such Rabi oscillations between long-lived states of Sr$_2$ molecules are described in Ref. \cite{mcguyer:1g}.
Figure \ref{fig4}(a) schematically illustrates another example of a coherent molecular process.
Here, two-color Raman transitions are driven between the two most weakly bound vibrational levels in ground-state $^{88}$Sr$_2$.
Multiple cycles of Rabi oscillations are observed for nearly 10 ms as shown in Figure \ref{fig4}(b), by monitoring either the initial (black circles) or the final (green triangles) state population using the detection methods described in Section \ref{sec:MoleculeSpectroscImaging}.
The $\sim100$ Hz spectral linewidth in Figure \ref{fig4}(c) was achieved by optimizing the probe pulse length and intensity.  It is limited both by collisions between the molecules within the 1D-lattice `pancakes' and by a weak state sensitivity of the optical lattice which maps the finite molecular temperature onto the transition line shape \cite{mcdonald:therm}.
The linewidths and coherence times can be further improved by actively stabilizing the lattice intensity, isolating fewer molecules per lattice site, and searching for `magic' trap conditions \cite{YeSci08}.

The ability to transfer population among weakly bound vibrational levels with a narrow spectral width allows highly precise measurements of ground-state binding energy differences, as given in Table \ref{tab:3}.  The experimental uncertainties are $\sim100$ Hz, including statistics and systematic effects, and can be significantly reduced in future measurements.
The dominant source of uncertainty in Table \ref{tab:3} comes from the kHz-level inaccuracy of the RF signal generator used in the experiment, which can readily be improved, particularly with the use of an optical frequency comb. Moreover, a bound-free two-color procedure
can be adopted to measure absolute ground-state binding energies using PD to the ground-state atomic continuum.

\section{Outlook}
\label{sec:Outlook}
The high-resolution optical lattice spectroscopy and coherent control demonstrated in this work set the stage for ultracold Sr$_2$ molecules to assume a prime role in precision metrology and fundamental science.
To date, these molecules have yielded insights into basic atom-molecule asymptotic physics \cite{mcguyer:1g,mcguyer:2013}.  The accessibility and coherent manipulation of ground-state vibrational levels can enable a self-normalizing molecular clock operating at frequencies of tens of terahertz \cite{ZelevinskyPRL08}.  Such a clock could improve laboratory limits on cosmological variations of the electron-to-proton mass ratio \cite{ChardonnetShelkovnikovPRL08_muStability} in a nearly model-independent way \cite{SchwerdtfegerBeloyPRA11_AlphaVarInSr2}.  Furthermore, the narrow optical molecular transitions may allow a higher sensitivity to quantum electrodynamical retardation effects than can be reached with ultracold alkali-metal dimers \cite{HuletMcAlexanderPRA96_Li2QED,WilliamsJonesEPL96_Na2QED}.
In addition, Sr$_2$ molecules in an optical lattice are a promising system to search for non-Newtonian, mass-dependent forces at the nanometer scale \cite{AdelbergerARNPS03}.  The abundance of spinless bosonic Sr isotopes suggests that fitting mass-dependent corrections to the ground-state Born-Oppenheimer interatomic potential \cite{PachuckiPuchalskiPRA10_SLithiumIonizPotential} will extract a highly competitive constraint.  The progress described here lays the foundation for these fruitful research directions.

\begin{table}
\caption{\label{tab:3}
Binding energy differences for X$^1\Sigma_g^+(v,J=0)$ molecules obtained by two-color Raman transitions.
Experimental uncertainties (``exp'') include statistical scatter and extrapolation to zero lattice and probe light powers, while calibration uncertainties (``cal'') include RF signal generator inaccuracy.
The values from this work are roughly three orders of magnitude more precise than previous values obtained by two-color PA \cite{killian:2008,reinaudi:2012}.}
\begin{indented}
\item[]\begin{tabular}{@{}l l l l}
\br
		& 		& \centre{2}{$E_b(v_1) - E_b(v_2)$ (MHz)} \\
\ns \ns
		& 		& \crule{2} \\
\ns
$v_1$	& $v_2$ 	& This work	& Previous work \cite{reinaudi:2012,killian:2008} \\
\mr
$-1$		& $-2$ 	& 1263.673582 $\pm$ (63)$_{\rm exp}$ $\pm$ $(320)_{\rm cal}$		& 1263.4(3) \\
$-2$		& $-3$ 	& 3710.255610 $\pm$ (170)$_{\rm exp}$ $\pm$ $(930)_{\rm cal}$	& 3710.5(2) \\
\br
\end{tabular}
\end{indented}
\end{table}

\ack
We are grateful to W. Skomorowski and R. Moszynski for providing {\it ab initio} theoretical data on molecular transition strengths, and to C. B. Osborn and G. Reinaudi for contributions to the experimental setup and to ultracold molecule spectroscopy at the early stages.  This work was partially supported by the ONR grant N00014-14-1-0802 and the NSF award PHY-1349725.  G. Z. I. acknowledges the NSF IGERT DGE-1069240.

\section*{References}

\providecommand{\newblock}{}

\end{document}